\documentclass[prl,aps,twocolumn,showpacs]{revtex4}
\usepackage[dvips]{graphicx}
\usepackage{dcolumn}
\newcommand{\be}{\begin{equation}}
\newcommand{\ee}{\end{equation}}
\newcommand{\bea}{\begin{eqnarray}}
\newcommand{\eea}{\end{eqnarray}}
\newcommand{\nn}{ \nonumber}
\newcommand{\ds}{\displaystyle}
\begin{document}
\topmargin=-20mm

\title{Fermi-liquid theory of the surface impedance of a metal in a normal magnetic field}

\author{Natalya A. Zimbovskaya }

\affiliation
{Department of Physics and Electronics, University of Puerto Rico at Humacao, Humacao, PR 00791 }

\begin{abstract}
 In this paper we present detailed theoretical analysis of the frequency and/or magnetic field dependence of the surface impedance of a metal at the anomalous skin effect. We calculate the surface impedance in the presence of a magnetic field directed along the normal to the metal surface. The effects of the Fermi-liquid interactions on the surface impedance are studied. It is shown that the cyclotron resonance in a normal magnetic field may be revealed {\it only and exclusively} in such metals whose Fermi surfaces include segments where its Gaussian curvature turns zero. The results could be applied to extract extra informations concerning local anomalies in the Fermi surface curvature in conventional and quasi-two-dimensional metals.
   \end{abstract}

\date{\today}
\maketitle

\section{I. Introduction}

Theoretical studies of the anomalous skin effect in metals were started long ago by Pippard \cite{1}, Reuter and Sondheimer \cite{2} and Dingle \cite{3}. Soon it became clear that the surface impedance of a metal under the anomalous skin effect contains informations concerning geometrical characteristics of the  Fermi surface (FS), especially its Gaussian curvature \cite{4,5}. When a metal is subject to both high frequency electromagnetic field $ \bf E $ and a constant and uniform magnetic field $ \bf B, $ the surface impedance depends on the magnitude of the magnetic field and its orientation with respect to the metal surface, so that some interesting effects may be manifested giving opportunities to extract extra informations on the electronic characteristics of metals \cite{6}, including the cyclotron resonance.

This resonance is displayed in the frequency/field dependence of the surface impedance while the incident electromagnetic field frequency $ \omega $ equals to $ n$-fold cyclotron frequency of electrons $ \Omega. $ There are two principal geometries providing the resonance features in the surface impedance to be revealed. At first, strong resonance occurs when the field $ \bf B $ is parallel to the surface of the metal, as predicted by Azbel and Kaner \cite{7} and repeatedly observed in experiments.
For such orientation, the electrons spiralling around the field may repeatedly return to the skin layer where they gain the energy from the electromagnetic wave. Also, the resonance can appear when the magnetic field is perpendicular to the surface of the metal \cite{8}. In this geometry some conducting electrons can stay inside the skin layer for a long while absorbing the energy. The cyclotron resonance in a normal magnetic field was viewed in semiconductors but it  was not expected to appear in good metals for their skin layers are too thin at high frequencies to give room to a sufficient number of electrons. However, this effect was observed in potassium \cite{9}, cadmium and zinc \cite{10,11} as well as in some layered organic metals \cite{12,13}.

Originally, the peak in the surface impedance in potassium was thought to arise due to the effect of the Fermi-liquid cyclotron wave \cite{9,14}. The latter is a transverse collective mode travelling in the electron Fermi-liquid along the magnetic field $ \bf B, $ as first predicted by Silin \cite{15}. Later, another explanation of the cyclotron resonance in potassium was offered in the paper by Lacueva and Overhauser \cite{16} assuming that the Fermi suface  of this metal includes some cylindrical pieces. This assumption was based on the results concerning the FS geometries of alkali metals which follow from the charge-density-wave theory \cite{17,18,19}.
So, while within the first approach the cyclotron resonance in a normal magnetic field in potassium is treated as a Fermi-liquid effect, the second approach consideres it as originating from the particular FS geometry.

To choose between the two hypotheses one must carry out a thorough calculation of the surface impedance taking into account Fermi-liquid effects and keeping all terms of the order or greater than $ \xi^{-3} $ in the expansions of the surface impedance components in the inverse powers of the anomaly parameter $ \xi $ (at the anomalous skin effect $ \xi \gg 1).$ This is a difficult and nontrivial computational task which is not yet completely executed. Currently, there exists a detailed analysis of the magnetic field dependence of the metal surface impedance at the anomalous skin effect completed by Kobelev and Silin accepting that the FS is a sphere \cite{20}. They did study Fermi-liquid effects on the surface impedance and arrived at the conclusion that a tiny resonance feature is to be displayed  at $ \omega = \omega_0 \ (\omega_0 $ is the limiting frequency of the Fermi-liquid cyclotron wave which in alkali metals is expected to take on value close to the electron cyclotron frequency $ \Omega.)$ However, the limitations of the employed model did not allow them to analyze the effects of the FS geometry. Also, in some other papers the analysis was concentrated on the effects of the FS geometry (see e.g. \cite{21,22,23,24}). It was shown that the cyclotron resonance in a metal in a normal magnetic field may occur due to some local anomalies in the FS curvature. Namely,
when the FS includes lines/points where its Gaussian curvature becomes zero, then the resonance features could be manifested in the frequency/field dependence of the surface impedance at $ \omega = \Omega. $ Obtained results agree with the above mentioned experiments \cite{9,10,11,12,13}.

Thus,  significant points are missing in the current theory. Possible effects of the electron-electron interactions are omitted from consideration in Refs. \cite{21,22,23,24}, whereas the analysis developed in Ref. \cite{20} lacks to consider the effects of the FS shape. The present paper is intended for clarifying  these points. Here, we calculate the surface impedance of a metal at the anomalous skin effect and in the presence of an external magnetic field taking into account both FS anisotropy and Fermi-liquid effects. As shown below, the Fermi-liquid cyclotron mode may influence the surface impedance in a manner similar to that obtained for a Fermi sphere \cite{20}, and we compare the corresponding features in the frequency/field dependence of the surface inpedance with those originating from the FS local geometry \cite{24}.

Such analysis is necessary to finally clarify the nature and origin of the cyclotron resonance in metals in the normal magnetic field. In particular, the obtained results may be useful in studies of electron characteristics of quasi-two-dimensional conducting materials (e.g. organic metals) which have attracted a great deal of interest in recent years. The outline of the paper is as follows. In Sections II, III, we consider the electron conductivity tensor taking into account Fermi-liquid interactions. In Sec. IV we calculate the surface impedance components. Finally, we discuss our results in Sec. V.

\section{II. Conductivity of the electron Fermi-liquid; Fermi-liquid cyclotron waves}

We consider a semi-infinite metal which fills the half-space $ z\leq0.$ The electromagnetic wave with the frequency $ \omega $ and the wave vector $ \bf q $ srikes the surface of the metal at zero angle of incidence, so $ {\bf q} = (0,0,q). $ We assume that the external magnetic field $ \bf B $ is directed perpendicularly to the surface.  We restrict our consideration with the case of an axially symmetric Fermi surface whose symmetry axis is parallel to the magnetic field.
Then the  response of the electron liquid of the metal to an electromagnetic disturbance  could be expressed in terms of the electron conductivity components $ \sigma_\pm (\omega,{\bf q}) = \sigma_{xx} (\omega,{\bf q} )\pm i \sigma_{yx} (\omega,{\bf q} ). $
The above restriction on the FS shape enables us to analytically calculate the conductivity components and to derive the explicit expressions for the latter presented below (see Eq. (15)). In general case, insurmounting difficulties arise in calculations of the electron conductivity when the effects of the Fermi-liquid interactions are included in consideration. The result for the electron conductivity (15) is  crucial in further studies of the Fermi-liquid effects in the surface impedance of a metal. Also, the recent analysis carried out in Ref. \cite{24} showed that no qualitative differences were revealed in the expressions for the principal terms of the surface impedance computed for the axially symmetric FSs and those not possessing such symmetry. This gives grounds to expect the currently employed model to catch the main features in the electronic response which remain exhibited when the FSs of generalized (non axially symmetric) shape are taken into consideration.

Within the phenomenological Fermi-liquid theory electron-electron interactions are represented by a self-consistent field affecting any single electron included in the electron liquid. Due to this field the electron energies $ E\bf (p) $ get renormalized, and the renormalization corrections depend on the electron position $ \bf r $ and  time $ t: $
\be 
 \Delta E = T r_{\sigma'} \int \frac{d^3 \bf p'}{(2 \pi \hbar)^3}\, F ({\bf p, \hat \sigma; p', \hat \sigma') \delta \rho (p', r',\hat\sigma',} t) .
  \ee
  Here, $ \delta \rho {\bf (p, r, \hat\sigma,} t) $ is the electron density matrix, $ \bf p $ is the electron quasimomentum, and $ \hat\sigma $ are spin Pauli matrices. The trace is taken over spin numbers $\sigma. $ The Fermi-liquid kernel included in Eq. (1) is known to have a form:
   \be 
 F ({\bf p, \hat\sigma; p', \hat\sigma') = \varphi (p,p')} + 4 \bf (\hat\sigma \hat\sigma') \psi (p,p')
  \ee
   For an axially symmetric FS the functions $ \varphi \bf (p,p')$ and $ \psi \bf (p,p')$  do not  vary under identical change in directions of projections $ \bf p_\perp $ and $ \bf p_\perp' $ of the electron quasimomenta $ \bf p $ and $\bf p'$ on a plane $ p_z = 0. $ These functions actually depend only on cosine of an angle $\theta$ between the vectors $\bf p_\perp $ and $\bf p_\perp'$ and on the longitudinal components of the quasimomenta $ p_z $ and $ p_z'$.

We  can separate out even and odd in cosine $\theta $ parts of the Fermi-liquid functions. Then function $ \varphi \bf (p,p')$ can be presented in the form \cite{25}
   \be
 \varphi({\bf p,p'}) = \varphi_0 (p_z,p_z', \cos\theta) + ({\bf p_\perp p_\perp'}) \varphi_1 (p_z,p_z',\cos\theta),
  \ee
  where $ \varphi_0,\varphi_1 $ are even functions of $ \cos\theta.$ Due to invariancy of the FS under  the replacement
$\bf p \to - p $ and $\bf p'\to -p',$ the functions $ \varphi_0 $ and $ \varphi_1 $  should not vary under simultaneous change in signs of $ p_z $ and $ p_z'.$ Using this, we can separate the functions $ \varphi_0,\varphi_1 $ into the parts which are even and odd in $ p_z, p_z',$ and to rewrite Eq. (3) as:
   \be
\varphi (p_z,p_z',\cos \theta) = \varphi_{00} + p_z p_z'\varphi_{01} + ({\bf p_\perp p_\perp'}) (\varphi_{10} + p_z p_z'\varphi_{11}).
  \ee
  The function $ \psi \bf (p,p')$ may be presented in the similar way. In the Eq. (4) the functions $ \varphi_{00},\varphi_{01}, \varphi_{10},\varphi_{11} $ are even in all their arguments,
namely: $ p_z,p_z'$ and $ \cos \theta .$ Each term in Eq. (4) corresponds to a particular part in the expansion of the Fermi-liquid functions in spherical harmonics
$ Y_{jm} (\Theta,\Phi) $ where the angles $ \Theta, \Phi $ determine the electron position on the Fermi sphere whose radius is $ p_F. $ For the function $ \varphi \bf (p,p') $ the expansion has the form:
  \be 
 \frac{2}{(2\pi\hbar)^3} \frac{p_F^2}{v_F} \varphi {\bf (p,p')} = \sum_{j=0}^\infty \sum_{|m|\leq j} \lambda_j Y_{jm} (\Theta,\Phi) Y_{j-m} (\Theta',\Phi').
    \ee
  Here, $ v_F $ is the Fermi velocity of electrons. Comparing Eqs. (4) and (5) we see that the function $ \varphi_{00} $ matchs that part of the expansion (5) which includes all terms with even values of both $ j $ and $m. $ The product $ p_z p_z'\varphi_{01} $ represents the sum of all terms in Eq. (5) with odd values of $j $ and even values of $ m. $ The third addend in Eq. (4) corresponds to the part of the expansion (5) containing terms with the odd values of both indices. Finally, those terms in the Eq. (5) labelled with even $ j $ and odd $ m $ are matched with the expression $ {\bf (p_\perp p_\perp')} p_z p_z'\varphi_{11}. $  As shown before \cite{15} the main Fermi-liquid effects in the response of the isotropic electron liquid to an external disturbance propagating along the magnetic field could be adequately analyzed while keeping the terms with $ j \leq 2 $ in the expansion (4). Omission of all terms with $ j > 2 $ from this expansion complies with the assumption that the functions $ \varphi_{00}, \varphi_{01}, \varphi_{10}, \varphi_{11} $ are constants. Here, we adopt the same approximation for these functions for an arbitrary axially symmetric FS.

In the following computations of the electron conductivity we start from the linearized transport equation for the nonequilibrium distribution function $ g {\bf (p,r,} t) = Tr_\sigma (\delta \rho {\bf ( p, r,\hat\sigma,} t). $ While considering a simple harmonic disturbance $ {\bf E = E}_{q\omega}\exp (i {\bf q\cdot r}- i \omega t), $ we may represent the coordinate and space dependencies of the distribution function $ g {\bf (p,r,}t ) $ in a similar way, namely:
 $ g {(p,r,}t) = g_{q\omega} \exp (i {\bf q r}- i\omega t). $ Then the linearized transport equation for the amplitude $ g_{q\omega} \bf (p) $ takes on the form:
  \be 
 \frac{\partial g^e_{q\omega}}{\partial \tilde t} + i{\bf q \cdot v} g_{q\omega}^e + \Big(\frac{1}{\tau} -i\omega\Big)g_{q\omega} + e\frac{\partial f_{\bf p}}{\partial E_{\bf p}} {\bf v E}_{q\omega} = 0.
  \ee
 Here, $f_{\bf p} $ is the Fermi distribution function for the electrons with energies $ E\bf (p), $ and $ {\bf v} = \partial E/\partial \bf p $ is the electrons velocity. The collision term in the Eq. (6) is written using the $ \tau $ approximation $(\tau $ is the electron scattering time) which is acceptable for high frequency disturbances $(\omega \tau \gg 1) $ considered in the present work. The derivative $ \partial g_{q\omega}^e/\partial \tilde t $ is to be taken over the variable $ \tilde t $ which has the meaning of time of the electron motion along the cyclotron orbit. The function $ g_{q\omega}^e \bf (p)$ introduced in the Eq. (6) is related to $ g_{q\omega} \bf (p)$  as follows:
  \be 
   g_{q\omega}^e {\bf (p)} = g_{q\omega} {\bf (p)} - \frac{\partial f_{\bf p}}{\partial E_{\bf p}} \sum_{\bf p} \varphi {\bf (p,p')} g_{q\omega} {\bf (p)} .
  \ee
  So, the difference between the distribution functions $g_{q\omega} {\bf (p)} $ and $g_{q\omega}^e {\bf (p)} $ originates from the Fermi-liquid interactions in the system of conduction electrons.

Substituting the approximation (4) for the Fermi-liquid function $ \varphi \bf (p,p') $ into Eq. (7), we can transform Eqs. (6), (7) into the system of linear equations for the averages:
  \be 
 I_1^\pm = \sum_{\bf p} g_{q\omega} {\bf (p)} p^\pm, \qquad
 I_2^\pm = \sum_{\bf p} g_{q\omega} {\bf (p)}p_z  p^\pm
  \ee
  where $ p^\pm = p_x \pm i p_y. $ The system appears to have the form:
  \bea 
 I_1^\pm \left \{1 + f_1 - \frac{f_1}{Q_0} \Phi_0^\pm \right \} - I_2^\pm \frac{f_2}{Q_2} \Phi_1^\pm = \frac{ie}{m_\perp} \Phi_0^\pm E_{q\omega}^\pm ,
 \nn\\
I_2^\pm \left \{1 + f_2 - \frac{f_2}{Q_2} \Phi_2^\pm \right \} - I_1^\pm \frac{f_1}{Q_0} \Phi_1^\pm = \frac{ie}{m_\perp} \Phi_1^\pm E_{q\omega}^\pm ,
  \eea
   Here, we introduce new Fermi-liquid parameters
  \bea
 && f_1 = \frac{2m_\perp}{(2 \pi \hbar)^3} \int p_\perp^2 \varphi_{10} dp_z, \nn \\
  && f_2 = \frac{2m_\perp}{(2 \pi \hbar)^3} \int p_\perp^2  p_z^2 \varphi_{11} dp_z.
    \eea
  We also use a notation:
   \bea
  \Phi_m^\pm &=& \int_{-1}^1 \frac{\overline a (x) x^m dx}{u\chi_\pm - \overline v (x)},
   \\
Q_m &=& \int_{-1}^1 \overline a(x) x^m dx,
  \\
  \rho &=& \frac{A(0)}{  \pi^2 m_\perp v_m p_m},
   \eea
  and
  \bea 
 \overline v (x) = v_z/v_m, \qquad &&
 \alpha_{1,2} = f_{1,2}/(1 + f_{1,2}),
  \nn\\
 x = p_z/p_m, \qquad && \overline a (x) = A (x)/ A(0), \nn \\
 u = \omega /q v_m, \qquad && \chi_\pm = 1 \pm \Omega/\omega + i/\omega \tau,  \nn
 \eea
  where $ p_m, v_m $ are the maximum values of the longitudinal components of the electron quasimomentum and velocity; $ A(x) $ is the cross-section area of the FS, $ \tau $ is the electron scattering time, $ m_\perp $ is the electron cyclotron mass, and $ E_{q\omega}^\pm = E_{q\omega}^x \pm i E_{q\omega}^y. $

Conductivity components $ \sigma_\pm (\omega,q) $ are determined by the expression:
  \be 
 \frac{e}{m_\perp} I_1^\pm = \sigma_\pm E_{q\omega}^\pm.
  \ee
 So, we solve the system (9), and we arrive at the following expressions for the circular components of the conductivity \cite{25}:
 \bea 
  \sigma_\pm & =&\frac{ie^2 p_m^2\rho}{4 \pi \hbar^3 q}
  \nn \\ &\times&
 \frac{\ds\bigg [\Phi_0^\pm \bigg(1 - \frac{\alpha_2 u}{Q_2} \Phi_2^\pm \bigg) + \frac{\alpha_2 u}{Q_2} (\Phi_1^\pm)^2 \bigg]}{\ds\bigg[ \bigg(1 - \frac{\alpha_1 u}{Q_2} \Phi_0^\pm \bigg) \bigg(1 - \frac{\alpha_2 u}{Q_2} \Phi_2^\pm \bigg) + \frac{\alpha_1 \alpha_2 u}{Q_0 Q_2} (\Phi_1^\pm)^2 \bigg]}.
 \nn \\
  \eea
   In calculation of the integrals $ \Phi_m^\pm (u) $ we have to take into account that because of the bilateral symmetry of the Fermi surface,  the longitudinal component of the velocity $ \overline v(x) $  and the cross-section area $ \overline a (x) $ are odd and  even functions of $ x, $ respectively. Combining contributions from symmetric segments of the FS we can carry out integration over the half-surface corresponding to positive  $ x. $

 When an external magnetic field is applied, electromagnetic waves may travel inside the metal. Here, we are interested in the transverse waves propagating along the magnetic field. The corresponding dispersion equation has the form:
  \be
  c^2 q^2 - 4 \pi i \omega \sigma_\pm (\omega,q ) = 0 .
  \ee
  When dealing with the electron Fermi-liquid, this equation for $``-"$ polarization  has  solutions corresponding to helicon waves and Fermi-liquid cyclotron waves first predicted by Silin for the isotropic electron liquid \cite{15}.

Considering Fermi-liquid cyclotron waves we may simplify the dispersion equation (16) by omitting the first term. Also, we can neglect corrections of the order of $ c^2 q^2/\omega_p^2 \ (\omega_p $ is the electron plasma frequency) in the expression for the conductivity. Then the Fermi-liquid parameter $ \alpha_1 $ falls out from the dispersion equation, and the latter takes on the form:
   \be 
 \Phi_2^- - \frac{(\Phi_1^-)^2}{\Phi_0^-} = \frac{Q_2}{u \alpha_2}.
  \ee
 Expanding the integrals $ \Phi_m^- $ in powers of $ u^{-1} $ and keeping terms of the order of $ u^{-2} $ we get the dispersion relation for the cyclotron mode at small $q\ (u\gg 1):$
  \be 
  \omega = \Omega (1 + f_2) \bigg[1 + \frac{\eta}{f_2} \bigg(\frac{q v_m}{\Omega} \bigg)^2 \bigg].
  \ee
 where:
  \be 
 \eta = \left[\int_{-1}^1 \overline a(x) \overline v^2 (x) x^2 dx - \frac{1}{Q_0} \left(\int_{-1}^1 \overline a (x)\overline v(x) x dx \right)^2 \right]\frac{1}{Q_2}.
 \ee

For an isotropic electron liquid $ \eta = 8/35, $ and the expression (18) coincides with the Silin's result.

\section{III. Conductivity of the electron Fermi-liquid at the large \large $ q$}

When $ u \ll 1 $ it is suitable to expand the transverse conductivity in powers of $ u. $ The expansion has the form \cite{22,24}:
  \be 
 \sigma_\pm (\omega,q) = \sigma_0 (q) (1 + \Lambda_1^\pm u + \Lambda_2^ \pm u^2 + \dots),
  \ee
  where
 \bea 
&& \sigma_0 = \frac{e^2 p_0^2}{4 \pi \hbar^3 q}, \\
&& p_0^2 = 2 \sum_j \bigg(1- \frac{1}{2}\delta_{j0} \bigg) \frac{1}{K_j (0)}.
   \eea

Summation over $ j $ in Eq. (22) is carried out over all effective
cross-sections on the FS where $ v_z = 0, $ and $ K_j(0) $ is the FS
curvature at the $j$-th cross-section. As for the coefficients $
\Lambda_n^\pm, $ the method of their computation is briefly outlined
in the Appendix 1 of the present work. The resulting expressions for
the first two coefficients are given below:
  \bea 
  \Lambda_1^\pm & =& - \frac{i g}{\pi} \left (a\chi_\pm + \frac{\pi^2}{g^2}\alpha_1 - b^2 \alpha_2\right),
 \\
   \Lambda_2^\pm &=& d\chi_\pm^2 - 2a \alpha_1 \chi_\pm - 2 \alpha_2 \frac{b}{s}g \chi_\pm \nn\\&&- 2 \alpha_1 \alpha_2 b^2 - \alpha_1^2 \frac{\pi^2}{g^2}.
  \eea
    Here, $ a,b,c,d,g $ are dimensionless constants whose values are
    determined with the FS shape. The explicit expressions for these
    constants are given in the Appendix 1. For a spherical FS we
    have: $ a=4,\ b =-4/3,\ d=g=s = 1. $

Finally, we remark that the expansion (20) gives the correct form for the conductivity only
when the FS does not include nearly cylindrical strips where its Guassian curvature becomes
zero. If such strips exist, the additional terms $\sigma_a^\pm (\omega, q) $ may appear in
the expressions for the conductivity components \cite{22,24}. These extra terms are
proportional to $ \sigma_0 (q) (u \chi_\pm)^\beta $ where $ -1 < \beta < 0. $ This parameter
 $ \beta $ describes the shape of the strip: the closer is the value of $ \beta $ to $ -1 ,$
 the closer is the strip shape to a pure cylinder. The occurence of this special term may
 bring significant changes to the expression for the surface impedance, and it may account
 for the cyclotron resonance in a normal magnetic field to be revealed.

\section{IV. Surface impedance}

We calculate the surface impedance of a semi-infinite metal in the presence of a normal magnetic field. Calculations are carried out under the anomalous skin effect conditions, when the electron mean free path $ l $ is greater than the skin depth $ \delta. $  We employ the model of axially symmetric FS in our calculations, so the surface impedance tensor is diagonalized in circular components.
The resulting expressions depend on the character of electrons scattering by the metal surface. Assuming a simplified model of the perfectly smooth metal surface providing the specular reflection of the electrons one  obtains \cite{2,3}:
  \be 
Z_\pm = \frac{8 i \omega}{c^2} \int_0^\infty \frac{dq}{ 4 \pi i \omega \sigma_\pm/c^2 - q^2},
  \ee
  where $ \sigma_\pm $ are the circular components of the transverse conductivity.

In realistic metal samples the surfaces always reveal some roughness, so certain electrons undergo diffuse scattering colliding with the latter. When the diffuse scattering predominates, the circular components of the surface impedance have the form \cite{26}:
  \be 
 Z_\pm = 4 \pi^2 i \omega \bigg/ \int_0^\infty \ln (1 - 4 \pi i \omega \sigma_\pm /c^2 q^2) d q.
 \ee
  Due to the high density of conduction electrons in good metals the skin depth $ \delta $
  may be very small. Assuming the electron density to be of the order of $10^{30} m^{-3}$,
   and the mean free path $l \sim 10^{-3} m$ (a clean metal), we estimate the spin depth
   at the disturbance frequency $ \omega \sim 10^9 s^{-1} $ as $ \delta \sim 10^{-6} m. $
   Therefore, at high frequencies $ \omega $ the skin effect in good metals becomes extremely
   anomalous so that $ \delta/l \sim 10^{-2} \div 10^{-3} $ or even smaller. Under these
   conditions electrons must move nearly in parallel with the metal surface to remain in
   the skin layer for a sufficiently long while. The effect of the surface roughness on
   such electrons is rather small. As shown before \cite{27}, the electrons reflection
   from the metal surface at the extremely anomalous skin effect may be treated as nearly
   specular, and  the corrections $ \Delta Z_\pm $ originating from the diffuse scattering
   have the order of $ \delta/l. $ In the present work we calculate the surface impedance
   keeping all terms no less in the order than $ \xi^{-3}\ (\xi = l/\omega\tau \delta $
   is the anomaly parameter). So, the corrections arising due to the surface rounghness
   may be neglected when $ \Delta Z_\pm \sim Z_0 \delta/l $ is smaller than $ Z_0 \xi^{-2} $
   where $ Z_0 \sim \omega \delta/c^2$ has the order of the main approximation of the
   surface impedance at the anomalous skin effect. Comparing the above estimations we
   arrive at the conclusion that $ \delta Z_\pm $ may be omitted when $ \omega \tau/\xi \gg 1. $
   This inequality could be satisfied at $ \omega \tau \sim 10^2,\ \xi \sim 10 $ which
   agrees with the conditions of the experiments reported in Refs. \cite{9,10,11}.
   On these grounds we carry out further calculations assuming the specular reflection
   of the electrons from the surface of the metal, and we start from the expression (25)
   for the surface impedance.

To proceed we turn to the integration over a new variable $ t = (q\delta)^{-1}.$
Then we can divide the integration range into two segments with different asymptotic
behaviors of conductivity:
   \bea 
&& Z_\pm = Z_1^\pm + Z_2^\pm,
  \nn\\ &&
Z_1^\pm = - \frac{8 i \omega}{c^2}\delta \int_0^\xi \frac{dt}{1 - i t^3
\overline \sigma_\pm (t/\xi)},
  \\&&
Z_2^\pm = - \frac{8 i \omega}{c^2}\delta \int_\xi^\infty \frac{dt}{1 - i t^3
\overline \sigma_\pm (t/\xi)}.
  \eea
  Here $\overline \sigma_\pm = \sigma_\pm/\sigma_0. $

At the  anomalous skin effect the surface impedance can be expanded
in the inverse powers of the anomaly parameter. The main terms in
this expansion originate from the addend $ Z_1^\pm. $ These terms
can be readily found by expanding the integrand in Eq. (27) in
powers of the parameter $ \epsilon_\pm $ given by:
  \be 
  \overline \sigma_\pm = 1 + \epsilon_\pm.
 \ee
 This parameter is small for $ \xi \gg 1. $ Now, we  calculate $ Z_1^\pm $
 using the expansion (20) for $  \sigma_\pm. $

Following the way proposed in the earlier work \cite{24} we arrive at the
result for the term $ Z_1^\pm: $
  \bea 
 Z_1^\pm &=& \frac{8\omega}{c^2} \delta \bigg[\frac{\pi}{3\sqrt 3}(1 - i\sqrt 3)
 + \frac{2\pi}{9\sqrt 3} \frac{\Lambda_1^\pm}{\xi} (i - \sqrt 3)
  \nn\\
&+&\frac{\Lambda_1^{\pm 2} - \Lambda_2^\pm}{\xi^2} \bigg(\ln\xi + \frac{i\pi}{3}
\bigg) - \frac{1}{2\xi^2}
 \nn\\
&+& \frac{\Lambda_2^\pm}{3 \xi^2} - \frac{\Lambda_1^{\pm 2}}{2 \xi^2}
 -\frac{i \Lambda_1^\pm}{\xi^2} \bigg] + \delta Z_1^\pm.
  \eea
 To arrive at the result (30) we used three first terms in
the expansion of $ \overline \sigma_\pm $ in powers of the small
parameter $ u $ and we kept the terms of the order of $ (t/\xi)^2 $
in the expression for $ \epsilon_\pm . $ Taking into account next
terms in the expansion of $ \overline \sigma_\pm $ in powers of $ u
$ or keeping next terms in the expansion of $ \epsilon_\pm $ we
obtain that corresponding integrals diverge. Therefore we cannot
expand the correction $ \delta Z_1^\pm$ in the inverse powers of the
anomaly parameter. To calculate this correction we carry out the
analytical continuation of the integrand in the Eq. (27) over the
first quadrant of the complex plane, and we compute the integral
over the path shown in the Fig. 1a. Then we obtain:
  \be 
 \delta Z_1^\pm = - \frac{8 \omega\delta}{c^2}Y_1^\pm + \delta Z_1'^\pm
  \ee
  where the first term corresponds to the integral over the segment
  of the imaginary axis and the second one originates from the integration
  over the circular arc. Introducing a new variable $ y = \xi/\mbox{Im}t $
   we arrive at the following expression:
  \be 
Y_1^\pm = \frac{1}{\xi^2} \int_1^\infty \bigg( \frac{y}{\overline
\sigma_\pm(i/y)} - y - i \Lambda_1^\pm + \frac{\Lambda_1^{\pm 2}
- \Lambda_2^\pm}{y} \bigg) dy.
  \ee
 Again, in calculation of the term $ Z_2^\pm $ we analytically
 continue the integrand in the Eq. (28), and we choose the integration
 path as shown in the Figure 1b.

\begin{figure}[t]
\begin{center}
\includegraphics[width=8.4cm,height=4.6cm]{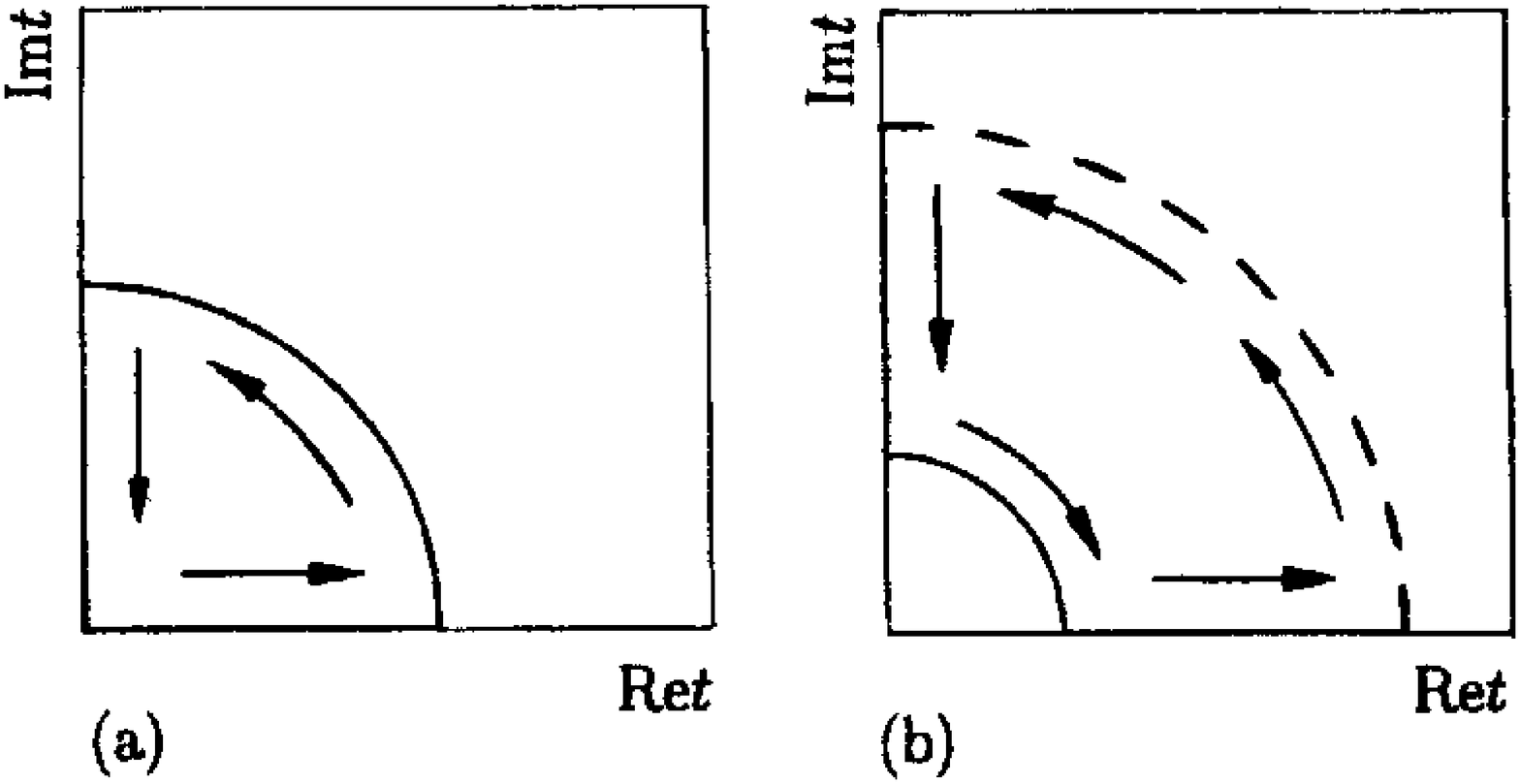}
\caption{Contours used for calculation of the quanties (a) $ Y_1 $ and (b) $ Z_2. $}
\label{rateI}
\end{center}
\end{figure}

Now, the integrand in the expression (28) has a pole originating
from the Fermi-liquid cyclotron wave (18). The pole appears in the
area bounded by the contour of integration  in the integrand of that
circular component of the surface impedance which corresponds to the
wave $(Z_2^- $ in our case). Accordingly, an extra contribution $
\delta Z_2^- $ from the residue  emerges in the expression for $
Z_2^-. $
 Also, the expression for $ Z_2^\pm $ includes the integral over the
 imaginary axis $( Y_2^\pm) $ and that over the circular arc $ (\delta Z_2'^\pm): $
  \be 
 Z_2^\pm = \delta Z_2^\pm - Y_2^\pm + \delta Z_2'^\pm
  \ee
  Here:
   \bea 
 \delta Z_2^+ &=& 0, \quad
 \delta Z_2^- = \frac{8 \omega \delta}{c^2} \alpha_2 ^2 W
 \sqrt{\left(\frac{\chi_-}{\alpha_2} - 1 \right) \frac{1}{\eta}},
 \\
 W &=& \frac{\pi^2}{2} \left(\int_{-1}^1 \overline a (x)
 \overline v (x) x dx \right)^2 \big / Q_0 Q_2\eta,
 \\
 Y_2^\pm &=& \frac{1}{\xi^2} \int_0^1 \frac{y}{\overline
 \sigma_\pm (i/y)} dy, \qquad y = \xi/\mbox{Im}t.
  \eea
 The last term in Eq. (33) is cancelled out with the term $\delta Z_1'^\pm ,$
 so we do not give its explicit expression. Combining these results,
  and using the expressions (23), (24) for $ \Lambda_{1,2}^\pm $ we get:
  \bea 
 Z_\pm = \frac{8 \omega \delta}{c^2} \bigg\{\frac{\pi}{3\sqrt 3}
 (1-i\sqrt 3)+ \frac{2 g}{9 \sqrt 3 \xi}
 \nn\\ \times
\Big(a\chi_\pm + \frac{\pi^2\alpha_1}{g^2} - b^2 \alpha_2 \Big)
 (1 + i\sqrt 3) + \frac{1}{\xi^2}\Big(\ln\xi + \frac{i\pi}{3} \Big)
\nn\\\times
\bigg[-\Big(\frac{g^2 a^2}{\pi^2} + d\Big) \chi_\pm^2
+ 2 \alpha_2 b g\Big(\frac{g a b}{\pi^2} + s \Big) \chi_\pm -
\frac{g^2 b^4 \alpha^2_2}{\pi^2} \bigg]
\nn\\+
 \frac{1}{\xi^2} \Big(\frac{1}{2} + \frac{d \chi_\pm^2}{3}
  +
\frac{g^2}{2 \pi^2}a^2 \chi_\pm ^2 - \frac{g}{\pi} a \chi_\pm \Big)
 \nn\\ -
\frac{\alpha_1}{\xi^2} \Big(\frac{a}{6} \chi_\pm + \frac{\pi}{g} \Big)  -
 \frac{\alpha_2 bg}{\xi^2} \Big(-\frac{2}{3s}\chi_\pm + \frac{b}{\pi} -
 \frac{g}{\pi^2} ab\chi_\pm \Big)
  \nn\\
- \frac{\pi^2 \alpha_1^2}{6 g^2\xi^2}  - \frac{b^2 \alpha_1 \alpha_2}{6 \xi^2}
 + \frac{b^4 g^2 \alpha_2^2}{2 \pi^2\xi^2}  - Y_1^\pm - Y_2^\pm \bigg\} +
 \delta Z_2^\pm. \nn\\
 \eea
 This result (37) is the correct and exact expression for the surface
 impedance circular components under the anomalous skin effect assuming
  that the FS of a metal everywhere possesses nonzero curvature. For a spherical
  FS $  W = 21/35 $ and $ \eta  = 8/35, $ so the expression (37) agrees with
 the corresponding result obtained for the particular case of isotropic
 Fermi-liquid \cite{20}.

Now, we must analyze the magnetic field dependence of the integrals
$ Y_1 $ and $ Y_2 $  included in the expression for the surface
impedance (37). These integrals have the form:
      \bea 
Y_1^\pm &=& \frac{1}{\xi^2} \int_1^\infty dy \bigg\{
\frac{y}{\overline \sigma_\pm (i/y)} - y
  \nn\\&+&
 \frac{g}{\pi} \left (a \chi_\pm +
\frac{\pi^2}{g^2} \alpha_1 - b^2 \alpha_2 \right)
 \nn\\ & +&
\frac{1}{y} \bigg[-
\left (\frac{a^2 g^2}{\pi^2} + d \right) \chi^2_\pm
 \nn\\ &+&
\left (\frac{a b g}{\pi^2} + s \right) \chi_\pm -
\frac{g^2 b^4 \alpha_2^2}{\pi^2} \bigg ] \bigg \},
   \\
Y_2^\pm &=& \frac{1}{\xi^2} \int_0^1 dy \frac{y}{\overline \sigma_\pm (i /y)}.
                      \eea
  The analysis is described in the Appendix 2. Assuming $ \omega \tau \to \infty,$
 we obtain the following asymptotic expressions for the integrals $ Y_{1,2} $ in the
 vicinity of the cyclotron resonance $ ( \chi_\pm \to 0):$
      \bea 
Y_1^\pm &=& \frac{1}{2 \xi^2}  - \frac{9}{\pi \xi^2}
\left (\frac{\pi^2}{g^2} \alpha_1 - b^2 \alpha_2 \right )
  + \frac{1}{\xi^2} \frac{g^2}{\pi^2} b^4 \alpha_2^2
 \nn\\&\times&
 \left [\ln |\alpha_2| -
\pi i \theta \left (\frac{\alpha_2}{\chi_\pm} \right )
- i \pi \left (1 - \theta (\chi_\pm) - \frac{i \pi}{2} \right ) \right ]
 \nn\\&+&
 o \left(\frac{\chi_\pm}{\xi^2} \right).
 \\
Y_2^\pm &=& - \frac{\chi_\pm}{\xi^2} \frac{\pi}{g} \alpha_1 +
\frac{\alpha_2^2}{\xi^2} + o \left ( \frac{\chi^3_\pm}{\alpha_2^3}
\right ).
                      \eea
  Here $ \theta (x) $ is a step function:
     \be 
\left \{ \begin{array}{l}
\theta (x) = 1, \qquad \ x\ge 0, \\
\theta (x) = 0,  \qquad \ x < 0.
\end{array} \right.
                    \ee

 It follows from Eqs. (40) and (41) that the integrals $ Y_{1,2} $
 do not include any terms showing resonance behavoir in the vicinity
 of the cyclotron frequency. So,
  the surface impedance of a metal whose Fermi surface everywhere has
   a finite and nonzero curvature in chosen geometry does not exhibit
   any resonance features corresponding to the cyclotron resonance in
   a normal magnetic field.
Not a single term in the expressions for the circular components
of the surface impedance reveals resonance features at $ \omega =\Omega $
or nearby, except $ \delta Z_2^-.$ The latter describes a singularity
in the surface impedance derivative at the extreme frequency of the
Fermi-liquid cyclotron wave $ \omega_0 = \Omega (1 + f_2 ). $ As for
the surface impedance itself, its frequency dependence reveals a kink
at $ \omega = \Omega. $ We evaluate the amplitude of the impedance
singularity due to the Fermi-liquid cyclotron wave under the conditions
typical for experimental observations of the cyclotron resonance
in the normal magnetic field $ (\omega\tau \sim 50, \ \xi^3
\sim 10^3\div 10^4). $ Under the assumption that the magnitude
of the Fermi-liquid parameter $ f_2 $ is not greater than $ 0.1, $
the magnitude of the resonance feature is about $ 10^{-4}\div 10^{-5} $
of the real part of the impedance. Resonance peaks at $ \omega = \Omega $
 may be revealed in the frequency/magnetic field dependence of the
 surface impedance when the FS includes nearly cylindrical segments.
 The resonance contribution to the surface impedance originates from
 the term $ \sigma_a^\pm (\omega,q) $ in the expression for the electron
 conductivity, and it is proportional to $ (\chi_\pm)^\beta. $ Considering
 these resonance features originating from the FS local geometry, it was
 shown that the resonance peak at $ \omega = \Omega $ has the magnitude
 of the order of $ 10^{-2} \div 10^{-3} $ of the real part of the impedance
 \cite{24}. This estimate was made assuming the same values for $ \omega\tau $
 and $ \xi $ as the previous ones. So, we see that resonance contribution
 to the surface impedance arising due to the FS local geometry predominates
 over that originating due to the Fermi-liquid effects.

Also, the height of the resonance peak produced by the contribution
from a zero curvature cross-section of the FS agrees with the
results of experiments \cite{9}, whereas the resonance feature
originating from the contribution of the Fermi-liquid cyclotron
wave is smaller by at least two orders in magnitude than these
experimental results. Even a stronger singularity in the impedance
derivative caused by this wave is too weak and cannot account
for the resonance features observed in cadmium and zink \cite{10,11}.

\section{V. Conclusion}

Basing on the obtained results we can make the following
conclusions. First, our analysis shows that the Fermi-liquid effects
may bring changes to the frequency/field dependence of the surface
impedance of a metal in a normal magnetic field but these changes
are minor. The resonance features at $ \omega = \Omega $ observed in
experiments on good metals cannot be attributed to these effects.
This provides a justification of the theory developed in Refs.
\cite{21,22,23,24}. Within the latter the cyclotron resonance in
metals in a normal magnetic field was attributed to the contribution
from zero curvature segments of the metal FS. Another, and the most
important result of the present analysis is that it demonstrates
impossibility for the resonance to be manifested in metals whose FSs
everywhere have nonzero Gaussian curvature. This directly follows
from Eqs. (37), (40), (41). All terms included in these expressions
describe a smooth field/frequency dependence of the surface
impedance components save  the addend  (34) describing the effect of
the Fermi-liquid cyclotron mode discussed before. So, within the
accepted model of the axially symmetric FS it is now proven that the
presence of zero curvature lines on the FS is the necessary
condition for the cyclotron resonance in a normal magnetic field to
be revealed in metals. Such proof is crucial to justify the theory
presented in Refs. \cite{21,22,23,24} and to make it conclusive.

This main result may be of practical use. If proven that the cyclotron
resonance in a normal magnetic field may be manifested only in those
metals whose FSs include zero curvature lines/points, then being actually
observed this effect can be treated as an evidence of these features in
the FS geometry of relevant metals. Therefore an extra information on
the electronic structure may be obtained basing on such observations.
This may appear to be especially useful in studies of electronic
structures of quasi-two-dimensional metals where very intensive
efforts are currently applied. Many such materials are high $T_C$
superconductors, and it is known that the FS curvature anomalies
could influence the transition temperature  $T_C$.

We got our results within the model of axially symmetric Fermi
surface assuming that the electron cyclotron mass $ m_\perp $ is
independent of $ p_z. $ However, the results may be easily generalized
to metals with cubic or hexagonal symmetries of the crystalline
lattice provided that the magnetic field is directed along a high-order
symmetry axis. Neglecting Fermi-liquid effects we can further
generalize our results but we do not believe this worthwhile, for
these generalizations in all probability will not bring qualitative
changes in the main results and conclusions presented above.

\section{Acknowledgments}

The author thanks G. M. Zimbovsky for help with the manuscript.
This work was supported  by NSF Advance program SBE-0123654, NSF-PREM 0353730,
and PR Space Grant NGTS/40091.

\appendix

\section{Appendix I}

In calculations of the integrals (11) we divide the FS in segments
supposing the one-to-one correspondence of $ p_z $ and $ v_z $  over
each segment. Then we change variables in these integrals and
rewrite them as follows:
  \bea 
 && \Phi^\pm_0 (u) = 2 u \chi_\pm \sum_i \int \frac{\overline a_i
  (\overline v) dx_i /d\overline v}{u^2 \chi_\pm^2 - \overline v^2}
  d \overline v, \\
  && \Phi^\pm_1 (u) = 2 \sum_i \int \frac{\overline a_i
  (\overline v)x_i (\overline v) dx_i /d\overline v}{u^2 \chi_\pm^2 - \overline v^2}
  d \overline v,  \\
  && \Phi^\pm_2 (u) = 2 u \chi_\pm \sum_i \int \frac{\overline a_i
  (\overline v)x_i^2 (\overline v) dx_i /d\overline v}{u^2 \chi_\pm^2
   - \overline v^2}  d \overline v.
  \eea

Summation over $ ``i" $ is carried out over the segments. Some
segments include effective cross-section $ \overline v =0. $
Cross-sectional areas have maxima/minima at $ \overline v = 0, $ so
we employ the following approximation for $ a_i (\overline v) $ near
$ \overline v = 0: $
   \be 
 a_i (\overline v) \approx a_i (0) - \tilde a_i (0) \overline v^2
  \ee
  where $ \tilde a_i (0) = \frac{1}{2} \nu_i |d^2 \overline a_i /d
 \overline v^2|_{\overline v=0}, $ and $ \nu_i $ is the sign
 factor. When $ a_i (\overline v ) $ has a maximum at $ \overline v
 = 0,\ \nu_i = 1. $ Otherwise, $ \nu_i = -1. $ Using the
 approximation (46) we present the contribution from the $j$-th
 segment including an effective cross-section to $ \Phi_0^\pm $ as
 follows:
   \bea 
&& 2 u \chi_\pm \int_{\overline v_{1j}}^{\overline v_{2j}}
\frac{[\overline a_j
 (\overline v)  - \overline a_j (0) + \tilde a_j (0)
 \overline v^2] dx_j/d \overline v}{u^2 \chi_\pm^2 - \overline v^2}d \overline v
   \nn\\
 &+& 2 u \chi_\pm \int_{\overline v_{1j}}^{\overline v_{2j}} \frac{[a_j (0) -
 \tilde a_j (0) \overline v^2] dx_j/d \overline v}{u^2 \chi_\pm^2 -
 \overline v^2}.
    \eea
 Here, $ \overline v_{1j} < 0 < \overline v_{2j}. $ The second term
 represents a singular part of the integral which gives the
 contribution from the vicinity of the effective cross-section $ \overline
 v = 0. $ This contribution equals:
  \be 
 - 2 u \chi_\pm \frac{d x_j}{d \overline v} \left
 (\frac{i\pi}{u\chi_\pm} + \frac{1}{|\overline v_{1j}|} + \frac{1}{|\overline
 v_{2j}|} \right) [\overline a_j (0) - \tilde a_j (0) (u \chi_\pm)^2]  .
   \ee
  Now, we expand the first (nonsingular) term in the expression (47)
  in series in powers of the parameter $ u \chi_\pm. $ This
  expansion takes on the form:
   \be 
  \sum_{n=0}^\infty {\cal M}_n (u \chi_\pm)^n
    \ee
  where
     \bea 
  {\cal M}_n = \int_0^{|\overline v_{2j}|} \frac{d x_j}{d \overline
  v} [\overline a_j (\overline v) - \overline a_j (0)]
  \frac{d \overline v}{\overline v^{2n+2}}
  \nn\\ -
  \int_0^{|\overline v_{1j}|} \frac{d x_j}{d \overline
  v} [\overline a_j (\overline v) - \overline a_j (0)]
  \frac{d \overline v}{\overline v^{2n+2}}.
  \eea
 Also, we expand in series in powers of $ u \chi_\pm $ all terms in the
 expression (43) originating from those segments which do not include
 effective cross-sections. Carrying out the summation over the
 segments and keeping all contributions no less than $(u \chi_\pm
 )^2$ we get the following approximation for $ \Phi_0^\pm: $
   \be 
 \Phi_0^\pm (u) = - \frac{i\pi p_0^2}{\rho p_m^2} \left[1 + \rho d (u
 \chi_\pm)^2 \right] -  u\chi_\pm.
   \ee
  Here,
  \bea 
 \! a \!&\! =\! & \!2 \sum_j \bigg [  \int_0^{|\overline v_{1j}|} \!
\frac{\gamma_j (\overline v)}{\overline v^2} d \overline v +\!
\int_0^{|\overline v_{2j}|} \! \frac{\gamma_j (\overline
v)}{\overline v^2} d \overline v + \frac{2}{\overline v} \bigg ]  (1
- \delta_{j0})
  \nn\\ &&+
2 \sum_k \bigg [  \int_0^{|\overline v_{1k}|}
\frac{ a_k (\overline v) \frac{d x_k}{d\overline v}}{\overline v^2} d\overline v
+\! \int_0^{|\overline v_{2k}|} \frac{ a_k (\overline v)
\frac{d x_k}{d\overline v}}{\overline v^2} d\overline v  \bigg ]
 \nn\\
&&+ 2 \bigg [  \int_0^{|\overline v_{0}|} \frac{\gamma_0 (\overline
v)}{\overline v^2} d \overline v - \frac{1}{\overline v} \bigg ]
\delta_{j0},
   \eea
           \bea 
&&  \gamma_j (\overline v) = \big [\overline a_j (\overline v) -
  \overline a_j (0) \big] \frac{d x_j}{d \overline v},
  \\
&&  d = \frac{2}{\pi^2\rho^2 p_m^2p_0^2} \sum_j \left(1- \frac{1}{2} \delta_{j0}
  \right)
   \frac{\nu_j}{\overline a_j (0) K_j^2 (0)}.
  \eea
 In Eq. (52), (54)  summation over $ ``k" $ is carried out over those segments of the
 FS which do not contain any effective  cross-sections, and summation over $ j $
 is carried out over effective cross-sections.

Proceeding in a similar way we obtain the following asymptotics for
the integrals $ \Phi_1^\pm: $
   \be 
  \Phi_1^\pm (u) = - i\pi \left(\frac{d x_j}{d\overline v} \right)^{-1}
  u \chi_\pm \delta_{j0} - b (u\chi_\pm)^2
    \ee
 where $ \delta_{j0} $ is the Kronecker delta symbol,
   \bea 
  b &=& 2 \int_0^{|\overline v|} \frac{d \overline v}{\overline v}
  \left( \overline a_0 (\overline v) x_0 (\overline v) \frac{dx_0}{d\overline
  v} \right)
    \nn\\ &+&
2 \sum_{i \neq 0} \bigg [\int_0^{|\overline v_{1i}|} \frac{d
\overline v}{\overline v} \overline a_i (\overline v) x_i (\overline
v) \frac{d x_i}{d \overline v}
   \nn\\&+&
 \int_0^{|\overline v_{2i}|} \frac{d
\overline v}{\overline v} \overline a_i (\overline v) x_i (\overline
v) \frac{d x_i}{d \overline v}.
  \eea
  All terms including the integrals $ \Phi_2^\pm $ in the
  expressions for the transverse conductivity components (15) are
  less than $ (u\chi_\pm)^2 $ in magnitude, so we drop them from
  consideration. Substituting Eqs. (51), (55) into (15), and
  introducing extra designations:
  \bea
 &&s =\big[\pi \rho p_0^2 p_m^2 K_0^2 (0) \big]^{-1},
 \nn\\
  &&g = \pi \rho p_m^2\big /p_0^2,
  \eea
 we arrive at the expressions (23), (24) for the coefficients $
 \Lambda_{1,2}^\pm. $

\section{Appendix II}

To study the magnetic field/frequency dependence of the integrals $
Y_1$ and $Y_2$ we make change of the variables in these integrals
introducing a new variable $ z\ (\chi_\pm z = y) $. Then we have:
    \bea 
Y_1^\pm &=& \frac{1}{2 \xi^2} (1 - \chi^2_\pm) - \frac{9}{\pi \xi^2}
\left (a \chi_\pm + \frac{\pi^2}{g^2} \alpha_1 - b^2 \alpha_2 \right
)
 \nn\\& +&
\frac{1}{2 \xi^2} \ln \chi_\pm \bigg [- \left (\frac{a^2 g^2}{\pi^2}
+ d \right) \chi^2_\pm
 \nn\\& +&
2 \alpha_2 b g \left (\frac{a b g}{\pi^2} + c \right) \chi_\pm +
\frac{g^2 b^4 \alpha_2^2}{\pi^2} \bigg ] - \frac{i \pi}{g}
\frac{\chi^2_\pm}{\xi} F_0
  \nn\\&
- &\frac{\pi}{g} \frac{\chi_\pm \alpha_2}{\xi^2} F_1 - \frac{i
\pi}{g} \frac{\alpha_2^2}{\xi^2} F_2 + \frac{\pi}{g} \frac{1}{\xi^2}
\frac{\alpha_2^3}{\chi_\pm} F_3^\pm.
                      \eea
 Here:
   \bea 
F_0 & =& \! \int_1^\infty \! \left \{\frac{z}{\Phi_0 (z)} - \frac{i
g}{\pi} z + \frac{i g^2 a}{\pi^2} - \frac{i g}{\pi z} \left
(\frac{a^2 g^2}{\pi^2} + d \right ) \right \} dz,
  \nn \\ \\
F_1 &=& \int_1^\infty \left \{ P(z) + \frac{g^2}{\pi^2} b^2 -
\frac{2}{z} b \frac{g^2}{\pi} \left (\frac{a b g}{\pi^2} + c \right
) \right \} dz,
     \\
F_2 &=& \int_1^\infty \left (\frac{P (z)}{z R (z)} - \frac{i}{z}
\frac{g^3}{\pi^3} b^4  \right ) d z,
                     \\
F_3^\pm &=& \int_1^\infty \frac{P (z)}{z^2 R(z)} \frac{d z}{1 - i
\alpha_2 / \chi_\pm z R (z)} .
                      \eea
 Besides we used notations:
      \bea 
P (z) &=& \frac{\Phi_1^2 (z)}{\Phi_0^2 (z)},
     \nn\\
R (z)& =& \frac{\Phi_0 (z)}{\Phi_0 (z) \Phi_2 (z) - \Phi_1^2 (z)}.
                      \eea
 Here $ \Phi_m (z) $ are the integrals $ \Phi^\pm_m$ defined by Eq. (11).
 The quantity $ u \chi_\pm $ in each integrand is replaced by $ i / z,$
  therefore $ \Phi_m (z) $ do not depend on the magnetic field.

The integral $ Y_2^\pm $ can be presented as:
    \bea 
Y_2^\pm & = & - \frac{\chi_\pm \pi}{g} \frac{\alpha_1}{\xi^2} -
\frac{i \pi}{g} \frac{\chi^2_\pm}{\xi^2} X_0 - \frac{\pi}{g}
\frac{\chi_\pm}{\xi^2} \alpha_2 X_1
  \nn\\&&-
\frac{i \pi}{g} \frac{\alpha_2^2}{\xi^2} X_2 +
\frac{\alpha_2^3}{\chi_\pm} \frac{\pi}{g} \frac{1}{\xi^2} X_3^\pm,
                   \\
X_0 &=& \int_0^1 \frac{z dz}{\Phi_0 (z)} ,
               \\
X_1 &=& \int_0^1 P(z) dz,
 \\
X_2 &=& \int_0^1 \frac{P(z)}{z R(z)} dz;
  \\
X_3^\pm &=& \int_0^1 \frac{P(z)}{z^2 R(z)} \frac{dz}{1 - i \alpha_2
/\chi_\pm z R(z)}.
                      \eea
  The magnetic field dependencies of $ Y_{1,2}^\pm $ are manifested through their dependencies of $ \chi_\pm. $ Corresponding factors are explicitly given in all terms included in Eqs. (58), (64) but $ F_3^\pm $ and $X_3^\pm .$ The integrals $ F_0, F_1, F_2 $ as well as $ X_0, X_1, X_2 $ do not depend on the magnetic field.
In further calculations we assume $ \omega \tau \to \infty, $ so $
\chi_\pm $ takes on real values. To compute the integral $ F_3^\pm $
we can write the expression for Mellin transform  with respect to
the variable $ \mu \ (\mu = |\alpha_2 /\chi|):$
      \bea 
\psi_\pm (k)& =& \! \int_0^\infty  \! d \mu \mu^{k-1} F_3^\pm (\mu)
\nn\\&=&
  \frac{\pi}{\sin \pi k} \exp \left [i \pi k \left (\theta \Big(\frac{\alpha_2}{\chi_\pm}\Big) + \frac{1}{2} \right ) \right ]
  \nn\\&\times&
\int_1^\infty P(z) (z R(z))^{k-2}dz.
                      \eea
   We can expand the integrand in series in the inverse powers of the variable $ z: $
      \be 
P(z) (z R(z))^{k-2} = z^{k-2} \sum \limits_{n=0}^\infty
\frac{\Delta_n(k - 2)}{z^n}.
                      \ee
 where $\Delta_n (k-2) $ are the coefficients in the expansion.
 Substituting this expansion into Eq. (69) we can rewrite the expression for $ \psi (k) $ in the form:
      \bea 
\psi (k) &=& \frac{\pi}{\sin \pi k} \exp \left [i \pi k \left
(\theta \Big(\frac{\alpha_2}{\chi_\pm} \Big) + \frac{1}{2} \right )
\right ] \sum \limits_{n=0}^\infty \frac{\Delta_n (k - 2)}{n - k +
1},
 \nn\\  k &< & n + 1.
                      \eea
 The inversed Mellin transformation gives:
      \bea 
F_3^\pm (\mu)& =& \frac{1}{2 \pi i} \int_{C - i \infty }^{C + i
\infty} \mu^{-k} \psi (k )dk
  \nn\\& =& -
\frac{i}{2} \sum_{n=0}^\infty \int_{C - i \infty }^{C + i \infty}
\exp \left [i \pi k \left (\theta
\Big(\frac{\alpha_2}{\chi_\pm}\Big) + \frac{1}{2} \right ) \right ]
  \nn\\&\times&
\frac{\mu^{-s}}{\sin \pi k} \frac{\Delta_n (k - 2)}{n - k + 1} ds .
                      \eea
 The integral included into Eq. (72) equals the sum of residues of the poles of the integrand arranged to the right from the line $ \mbox{Re} \,k = C \  (0 < C < n + 1). $ The contribution from the simple pole corresponding to $ k = 1, \ n \ne 0 $ is:
      \be 
\frac{i \chi_\pm}{\alpha_2^2} \int_1^\infty \left [\frac{P(z)}{z
R(z)} - \frac{\Delta_0 (-1)}{z} \right ] dz.
                      \ee
 Using the expressions (63) for the functions $ P(z) $ and $ R (z) $ and the asymptotics  for the integrals $ \Phi_0 $ and $ \Phi_1 $ at $ u \ll 1 $ (carrying this out we have to replace $ u \chi_\pm $ by $ i / z$ in the integrands) we obtain:
      \be 
\Delta_0(-1) = \frac{i g^3}{\pi^3} b^4.
                      \ee
 Thus the desired residue equals  $ i \pi \alpha_2^2 F_2 / g \xi^2.$

The contribution from the double pole corresponding to $ k = 1, \ n
= 0 $ is:
      \be 
\frac{g^3}{\pi^3} \frac{\chi_\pm}{\alpha_2} b^4 \left [\ln \left
|\frac{\alpha_2}{\chi_\pm} \right | - i \pi \theta \left
(\frac{\alpha_2}{\chi_\pm}\right ) - i \frac{\pi}{2} \right ].
                     \ee
 We also have to take into account the contribution from the simple pole $ (k = 2, \ n \ne 1) $ which equals:
      \be 
\frac{\chi^2_\pm}{\alpha_2^2} \int_1^\infty \left [P(z) - \Delta_0
(0) - \frac{\Delta_1 (0)}{z} \right ] dz =
\frac{\chi^2_\pm}{\alpha_2^2} F_1 .
                      \ee
 Here we used the expressions for $ \Delta_0 (0) $  and $ \Delta_1 (0):$
   \be 
 {\Delta_0 (0) = - \frac{g^2}{\pi^2}b^2},\qquad  \Delta_1
(0) = \frac{2 b g^2}{\pi} \left (\frac{a b g}{\pi^2} + s \right ).
  \ee
 Finally, we have  to consider the double pole $ (k = 2; \ n = 1), $ which gives the contribution:
      \be 
\frac{2 b g^2}{\pi} \left (\frac{a b g}{\pi^2} + c \right )
\frac{\chi^2_\pm}{\alpha_2^2} \left \{2 \pi i \theta \left
(\frac{\alpha_2}{\chi_\pm} \right ) + i \pi - \ln \left
|\frac{\alpha_2}{\chi_\pm} \right | \right \},
                      \ee
 and the contributions from the simple pole at $  k = 3;\ n \ne 2 $ and the double pole at $ k = 3, \ n = 2, $ as well. The simple pole gives us the following insertion:
     \bea 
&& - \frac{i \chi^3_\pm}{\alpha_2^3} \int_1^\infty dz \left
(\frac{z}{\Phi_0 (z)} + z \Delta_0 (1) + \Delta_1 (1) +
\frac{\Delta_2 (1)}{z} \right ) \nn\\&& = - i
\frac{\chi^3_\pm}{\alpha_2^3} F_0,
                      \eea
 and the contribution from the double pole equals:
     \be 
- \frac{\chi^2_\pm}{\xi^2}\left (\frac{a^2 g^2}{\pi^2} + d \right )
\left \{3 \pi i \theta \left (\frac{\alpha_2}{\chi_\pm} \right ) + 3
\pi \frac{i}{2} - \ln \left |\frac{\alpha_2}{\chi_\pm} \right |
\right \}.
                      \ee
 We do not account for  poles corresponding to larger values of $ k $ because their contributions are smaller in the order of magnitude than $\chi^2_\pm. $

So, we get:
 \bea 
Y_1^\pm &=& \frac{1}{2 \xi^2}  - \frac{9}{\pi \xi^2}
\left (\frac{\pi^2}{g^2} \alpha_1 - b^2 \alpha_2 \right )
  + \frac{1}{\xi^2} \frac{g^2}{\pi^2} b^4 \alpha_2^2
 \nn\\&\times&
 \left [\ln |\alpha_2| -
\pi i \theta \left (\frac{\alpha_2}{\chi_\pm} \right )
- i \pi \left (1 - \theta (\chi_\pm) - \frac{i \pi}{2} \right ) \right ]
 \nn\\&+&
 o \left(\frac{\chi_\pm}{\xi^2} \right).
                     \eea
 Also, we expand $X_3^\pm $ in powers of the small parametr $\chi_\pm /\alpha_2 $ and we obtain:
 \be 
X_3^\pm = \frac{i \chi_\pm}{\alpha_2} X_2 +
\frac{\chi^2_\pm}{\alpha_2^2} X_1 + \frac{\chi^3_\pm}{\alpha_2^3}
X_0 + o \left (\frac{\chi^4_\pm}{\alpha_2^4} \right ).
  \ee
 Substituting this result into Eq. (64) we arrive at the expression:
 \be 
Y_2^\pm = - \frac{\chi_\pm}{\xi^2} \frac{\pi}{g} \alpha_1 +
\frac{\alpha_2^2}{\xi^2} + o \left ( \frac{\chi^3_\pm}{\alpha_2^3}
\right ).
                      \ee

\end{document}